\documentclass{ws-procs9x6}

\begin{document}

\title{Light Sensor Candidates for the \\
Cherenkov Telescope Array}

\author{M. L. Kn\"otig$^*$, R. Mirzoyan, M. Kurz, J. Hose, E.
Lorenz, T. Schweizer, M. Teshima}
\address{Max-Planck-Institut f\"ur Physik, 80805 M\"unchen, Germany \\
$^*$E-mail: mknoetig@mpp.mpg.de}

\author{P. Buzhan, E. Popova}
\address{Moscow Engineering and Physics Institute, 115409 Moscow, Russia}

\author{J. Bolmont, J.-P. Tavernet, P. Vincent} 
\address{LPNHE, Universit\'e Pierre et Marie Curie Paris 6, 75252 Paris Cedex 5, France}

\author{M. Shayduk}
\address{Deutsches Elektronen-Synchrotron (DESY), 15738 Zeuthen, Germany
}

\author{On behalf of the\\ CTA Focal Plane Instrumentation WP}

\begin{abstract}
We report on the characterization of candidate light sensors for use in the next-generation Imaging Atmospheric Cherenkov Telescope project called Cherenkov Telescope Array, a major astro-particle physics project of about 100 telescopes that is currently in the prototyping phase. Our goal is to develop with the manufacturers the best possible light sensors (highest photon detection efficiency, lowest crosstalk and afterpulsing). The cameras of those telescopes will be based on classical super-bi-alkali Photomultiplier tubes but also Silicon Photomultipliers are candidate light sensors. A full characterisation of selected sensors was done. We are working in close contact with several manufacturers, giving them feedback and suggesting improvements.
\end{abstract}


\bodymatter
\section{Introduction}\label{aba:sec1}
The atmosphere is opaque for cosmic rays. But when particles with very high energy (VHE) of at least some tens of GeV hit the atmosphere they produce extended showers of secondary particles. These emit blueish Cherenkov light flashes due to their high speed, exceeding that of light in the atmosphere. The faint flashes can be collected by a telescope and projected onto it's fine pixelised camera. This method is called Imaging Atmospheric Cherenkov Telescope (IACT) technique and it is currently the most successful one with over 120 sources of VHE gamma-rays discovered \cite{ TevCat}. The camera consists of hundreds of ultra fast and highly sensitive light sensors, fast enough to follow the development of the shower\cite{ Ostankov01}. All the current telescopes use Photomultiplier Tubes (PMT) --- with the exception of the FACT telescope\footnote{Their approach is to use Silicon Photomultipliers (SiPM) from Hamamatsu (MPPC)\cite{ Anderhub11}}.

\section{The Cherenkov Telescope Array}
The Cherenkov Telescope Array (CTA) is the major project for the next generation ground based VHE gamma-ray astronomy. Current systems of IACTs use at most four telescopes. The plan is to build an array of about 100 telescopes of three different sizes (large $\sim$23m, middle $\sim$12m and small $\sim$6m) that will provide a ten times higher sensitivity compared to current systems. The camera of each telescope will comprise $\sim$2000 of ultra-fast PMTs which means a total need of $\sim 150000$ PMTs.

\begin{table}
\tbl{The FPI sensor wish list --- a selection of parameters in comparison with a target PMT}
{\begin{tabular}{@{}cccc@{}}\toprule
Parameter & Range Specification & Hamamatsu R11920-100 \\\colrule
Spectral Sensitivity Range \hphantom{00}&\hphantom{0}290 - 600 nm & 300 - 650 nm\\
Peak Quantum Efficiency\hphantom{00}&\hphantom{0}35\% & (35.6$\pm$1.7)\% \\
Average QE over Cherenkov Spectrum\hphantom{0}& $>21\%$ & (22.8$\pm$1.0)\%\\
Afterpulsing at 4 ph.e. Threshold & $<0.02\%$ & $\simeq$0.03\%\\
Transit Time Spread, single ph.e, FWHM & $<1.3$ ns & (1.3$\pm$0.1) ns\\
Collection Efficiency 1.st Dynode & 96\% & $\simeq$93\%\\
\botrule
\end{tabular}}
\label{aba:tbl1}
\end{table}

\section{Photomultiplier Tubes}
\begin{figure}
\begin{center}
\begin{tabular}{cc}
\psfig{file=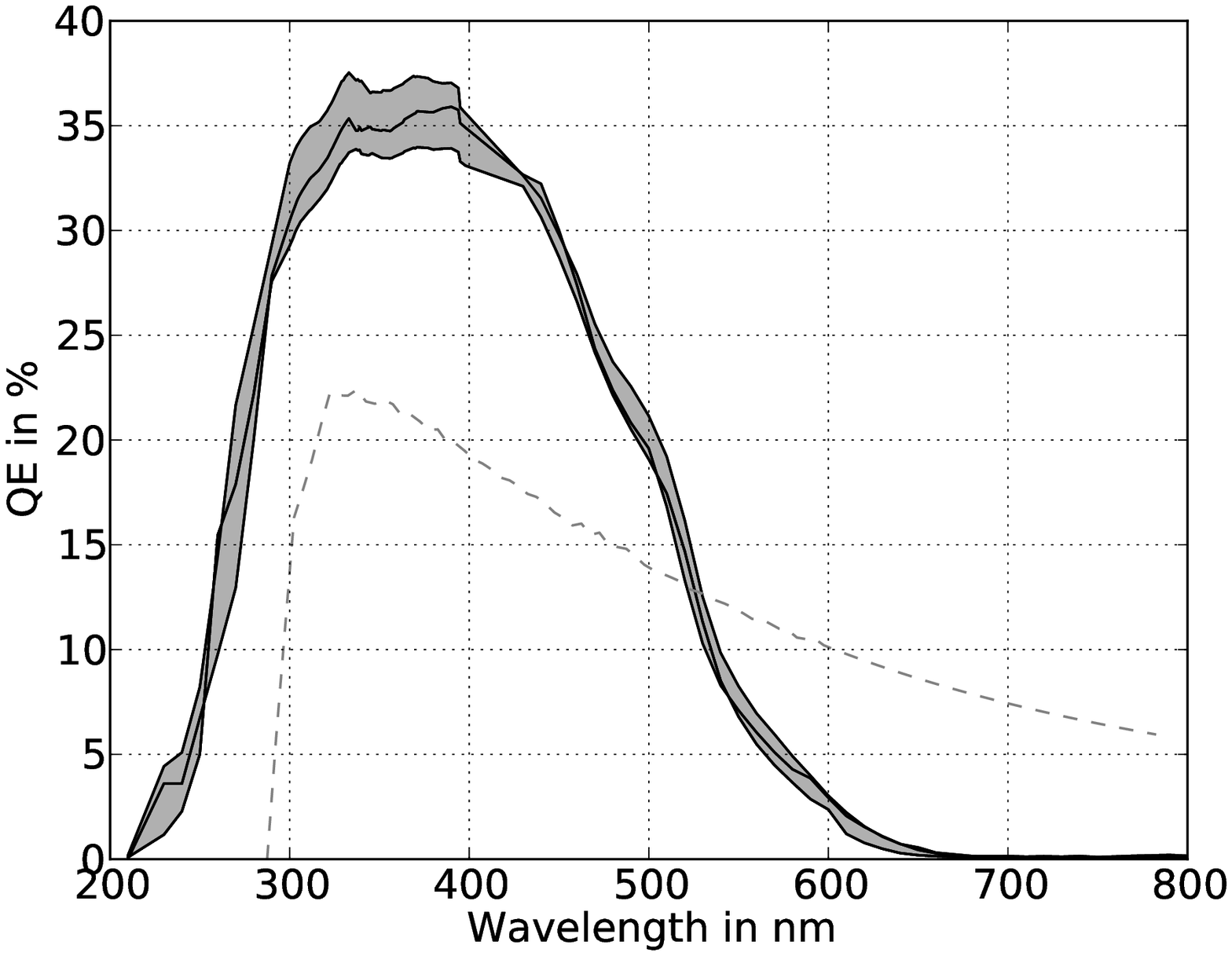,width=0.5\linewidth}&
\psfig{file=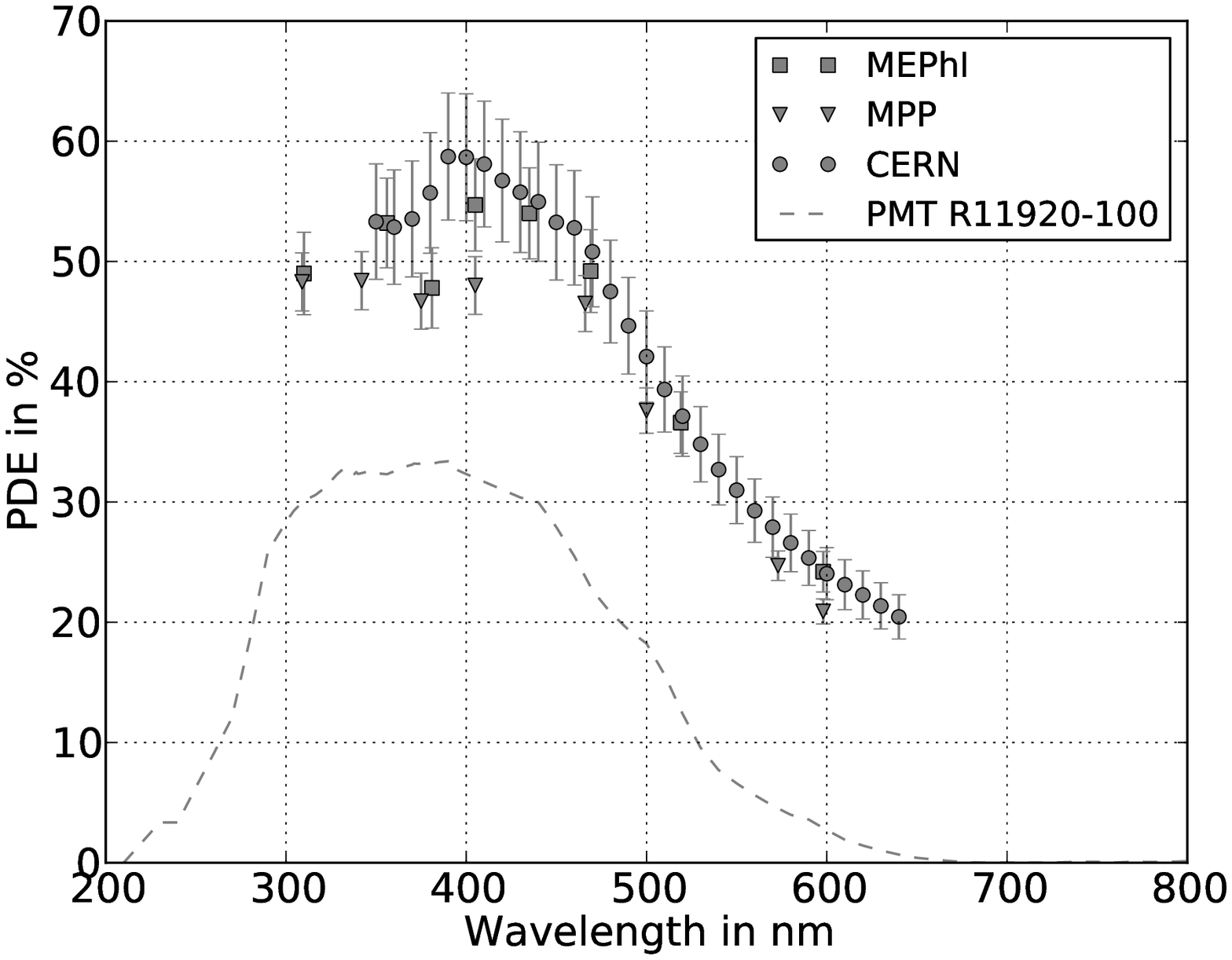,width=0.5\linewidth}
\end{tabular}
\end{center}
\caption{Left: Quantum efficiency of three R11920-100 samples over wavelength. The dashed line is the differential Cherenkov light spectrum of a 100GeV photon hitting the atmosphere at zenith angle = 0; Right: Photo detection efficiency of the MEPhI SiPM 100b 1x1mm structure compared to the R11920-100 (dashed). Circles: Y. Musienko, CERN; Squares: P. Buzhan, MEPhI; Triangles: M. Kn\"otig, MPP}
\label{aba:fig2}
\end{figure}

A development program was started with Hamamatsu K.K. and Electron Tubes Enterprises Ltd and a full characterisation of selected PMT samples was done. This combined measurements of quantum efficiency, afterpulses, single photo electron response, transit time spread and light emission. After two years the development program resulted in the PMT candidate R11920-100 from Hamamatsu which combines the cathode of the Hamamatsu R9420 --- providing very high quantum efficiency --- and the dynode structure of the Hamamatsu R8619 --- providing very low afterpulsing. In \tref{aba:tbl1} we show a comparison between some selected parameters from the wish list and the measurement results for three samples of the R11920-100. The quantum efficiency in particular is high and has its maximum at about 340--370 nm with an average peak quantum efficiency of (35.6$\pm$1.7)\%, as can be seen in \fref{aba:fig2}.

\begin{figure}
\begin{center}
\psfig{file=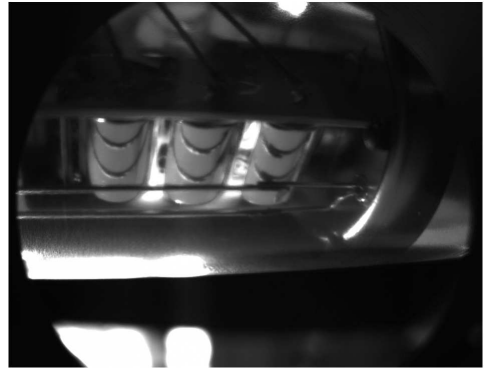,width=2.5in}
\end{center}
\caption{Overlay of an optical and a light emission picture showing the R11920-100 dynode structure. A pulsed laser is shooting at the cathode on the right at a rate of 20MHz and 440nm. The voltage applied is 1.1kV and the emission is integrated over 1s. One can see the light emission between dynodes two, four and six.}
\label{aba:fig3}
\end{figure}

PMTs show weak emission of light from the dynodes \cite{ Krall67}. For a better understanding of the R11920-100, we investigated and measured this(\Fref{aba:fig3}). We assume that this light, after many reflections, could arrive at the cathode and produce light-induced afterpulses. These shall become visible only after few tens of nanoseconds after the laser pulse illuminates the cathode. Such information is communicated to manufacturers and should help to improve the PMT, in this case the light-induced afterpulsing. We have a new setup with a fast gated image intensifier of the Hamamatsu C9546 series which will allow us to measure coincidences with a delayed pulse.

Hamamatsu and Electron Tubes have both worked to increase the collection efficiency (CE) of their PMTs. The next generation prototype from Hamamatsu reaches a CE of 96\% due to changes in the input window curvature and stabilised cathode-to-first-dynode voltage.

The main specifications (\Tref{aba:tbl1}) of CTA light sensors are close to being fulfilled by the currently developed PMTs. We are optimistic that they all will be met before the the beginning of the main construction phase.

\begin{figure}
\begin{center}
\psfig{file=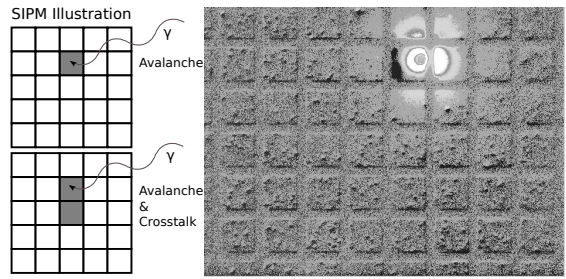,width=2.9in}
\end{center}
\caption{Left: Illustration of a crosstalk event in a SiPM, artificially increasing the signal; Right: Emission morphology from the SiPM surface, type 100b 1x1mm MEPhI at 38V, the grey spot marks the focused laser}
\label{aba:fig4}
\end{figure}

\section{Silicon Photomultipliers}
A Silicon Photomultiplier, known as SiPM but also as MPPC and G-APD array is a novel solid state photo sensor with a potential of 2-3 times higher photon detection efficiency (PDE) compared to classical PMTs. They consist of a matrix of small avalanche photo diodes with a common anode and are operated in Geiger mode. To become useful for CTA the PDE, the crosstalk between cells (\Fref{aba:fig4}) and the dark rate need to improve. In contrast to the PMT's afterpulses, reduction of the crosstalk is still the biggest challenge for the application of SiPM in the CTA project as it reduces the amplitude resolution and requires a higher trigger threshold\cite{ Buzhan09}.

We present the measurements of the recent prototype SiPM from Moscow Engineering and Physics Institute (MEPhI) in cooperation with Excelitas of the p-on-n type. As three independent measurements show it has a high and flat plateau with PDE(400 nm) $\approx50\%$ (\Fref{aba:fig2}) and good sensitivity in the ultraviolet region. The crosstalk effect was suppressed by introducing trenches between the cells, a second p-n junction and by ion implantation and is about 5\% at 14\% relative overvoltage, defined as $\tfrac{\Delta U}{U_{applied}}$ where $\Delta U = U_{applied} - U_{breakdown}$.

 We have developed a new method for imaging the crosstalk morphology. The idea is to shoot with a laser onto the surface of a selected cell and to count the number of photons emitted by the avalanches\cite{ Mirzoyan09} in the neighbouring cells. Interestingly, as one can see in \fref{aba:fig4}, the avalanches emit light only locally around the focused laser spot. Our comparison of this method with the classical one shows very good agreement.

\section{Conclusions}
The PMT development for CTA is in progress and already now has almost achieved the specified targets. During the next two years further optimization of PMTs from both Hamamatsu and Electron Tubes Enterprises is foreseen.

SiPM development is ongoing. We consider their use interesting in many applications when the PDE of SiPM will become 1.5-2 times higher than that of PMTs at comparable costs. New MEPhI SiPMs are getting there with peak PDE $\simeq$ 50\% at $P_{crosstalk}$ $\simeq5\%$. Soon these are going to become a commercial product. In a couple of years this type of sensor could become a serious alternative to the PMT in CTA.

\bibliographystyle{ws-procs9x6}
\bibliography{ws-pro-sample}

\end{document}